\documentstyle[12pt]{article}

\renewcommand{\theequation}{\thesection\arabic{equation}}

\begin{document}
\begin{center}
\title{\Large\bf Plane torsion waves in quadratic gravitational theories}
\author{O. V. Babourova\thanks{E-mail:baburova.physics@mpgu.msk.su},
        B. N. Frolov\thanks{E-mail:frolovbn.physics@mpgu.msk.su},
        E.A. Klimova\thanks{E-mail:eklimova@aha.ru}
}
\maketitle
{\it Department of Mathematics, Moscow State Pedagogical University,\\
     Krasnoprudnaya 14, Moscow 107140, Russia}
\end{center}
\begin{abstract}
{\small
The definition of the Riemann--Cartan space of the plane wave type is given.
The condition under which the torsion plane waves exist is found. It is
expressed in the form of the restriction imposed on the coupling constants of
the 10-parametric quadratic gravitational Lagrangian. In the mathematical
appendix the formula for commutator of the variation operator and Hodge
operator is proved. This formula is applied for the variational procedure
when the gravitational field equations are obtained in terms of the exterior
differential forms.
}
\end{abstract}

\section{\bf Introduction}
\setcounter{equation}{0}
\par
Recently the more attention is payed to the investigation of exact solutions
of field equations in the spaces with the geometrical structure which is
more complicated  that  the  Riemann  structure  of General Relativity.  The
problem of the wavelike solutions is on the special place here since this
problem is closely connected with the experimental investigation of
gravitational waves. In \cite{Ad} the gravitational waves of a metric and
torsion are considered in the theory with the Lagrangian which represents the
sum of the linear Einstein--Cartan Lagrangian, one of the six possible
quadratic in curvature terms and all the possible terms quadratic in torsion.
In \cite{Ch-Ch} the torsion waves on the flat space background are described.
The Lagrangian considered there is quadratic in curvature. In \cite{Sin} the
authors investigate the plane waves in the framework of the theory based on
the Lagrangian quadratic in curvature and torsion without linear term.
Ref. \cite{Zit} is devoted to the investigation of the waves of the torsion
2-form of the algebraic special N-type.
\par
In this article we consider the gravitational theory based on the general
quadratic Lagrangian in the Riemann--Cartan space-time $U_4$. The main
purpose is to investigate the problem of existance of the plane torsion waves
in $U_4$. We also want to clearify the role of each irreducible part of the
torsion propagating in the form of the plane wave. Preceding version of this
work can be found in Ref. \cite{trud}.
\par
     The modern Poincar\'{e} gauge gravitational theory essentially uses
the non-linear in curvature and torsion Lagrangians \cite{Fr1}--\cite{Ob-PZ}
(see also Refs. \cite{Iv-Sar}--\cite{He:pr} and references therein).
The use of the quadratic Lagrangians in the theory of gravitation is also
stimulated by the attempt to construct the renormalized theory of gravitation
in the Riemann--Cartan space-time \cite{Ya}, \cite{IvPrSar}. The most of the
quadratic gravitational theories in the Riemann--Cartan space-time can be
described as the particular cases of the 10-parametric Lagrangian, described
in Ref. \cite{Hay}. This Lagrangian is constructed as the sum of the
Einstein--Cartan linear Lagrangian and all the terms quadratic in the
irreducible pieces of curvature and torsion.

\section{\bf Field equations for the general quadratic Lagrangian}
\setcounter{equation}{0}
     The Riemann--Cartan space-time $U_4$ is the 4-dimentional oriented
differential manifold ${\cal M}$ with a metric $g_{ab}$ ($a,b = 0,1,2,3$)
(the metric has the Index $1$), the volume 4-form $\eta$, the linear
metric-compatible connection 1-form $\Gamma^{a}\!_{b}$, the curvature 2-form
${\cal R}^{a} \!_{b} = \frac{1}{2} R^a\!_{bcd}{\theta }^{c}\wedge
{\theta}^{d}$ and the torsion 2-form ${\cal T}^{a} = \frac{1}{2}T^{a}\!_{bc}
\theta^{b}\wedge\theta^c$. Here $\theta^a$ ($a = 0,1,2,3,$) represent the
cobasis of 1-forms in $U_4$ ($\wedge$ is the operation of the exterior
product). We use the local anholonomic vector basis $\bar{e}_b$ with
$\bar{e}_b\rfloor\theta^a = \delta^{a}_{b}$, where $\rfloor$ is the operation
of the interior product, and $g_{ab} = g(\bar{e}_a,\!\bar{e}_b) =  const$.
\par
     In $U_4$ a connection is compatible with a metric in the sence that the
exterior covariant differential $D = d + \Gamma\wedge\ldots$ ($d$ is the
operator of the exterior differential) of the metric tensor is equal to zero,
\begin{equation}
D g_{ab} = dg_{ab} - \Gamma^{c}\!_{a}g_{cb} - \Gamma^{c}\!_{b}g_{ac} =
- 2\Gamma_{(ab)} = 0\;. \label{eq:sovm}
\end{equation}
The condition (\ref{eq:sovm}) represents the constraint imposed on the
connection 1-form $\Gamma^{a}\!_{b}$. This constraint can be resolved
explicitly by using the connection 1-form which satisfies to the condition
$\Gamma_{ab} = - \Gamma_{ba}$.
\par
     It is convinient to use the  fields of 3-forms $\eta_a$, 2-forms
$\eta_{ab}$, 1-forms $\eta_{abc}$ and 0-forms $\eta_{abcd}$ , determined as
follows \cite{Tr},
\begin{eqnarray}
&&\eta_a = \bar{e}_a\rfloor \eta = *\theta_a \; ,  \qquad \eta_{ab} =
\bar{e}_b\rfloor \eta_a = *(\theta_a\wedge \theta_b) \;, \label{eq:eta1} \\
&& \eta_{abc} = \bar{e}_c\rfloor \eta_{ab} = *(\theta_a
\wedge \theta_b\wedge \theta_c) \; , \qquad
\eta_{abcd} = \bar{e}_d\rfloor \eta_{abc} = *(\theta_a
\wedge \theta_b\wedge \theta_c\wedge \theta_d) \; ,  \label{eq:eta2} \\
&&\theta^a \wedge \eta_b = \delta^{a}_{b}\eta \; , \qquad
\theta^a \wedge \eta_{bc} = -2\delta^{a}_{ [b}\eta_{c]} \; , \label{eq:eta3} \\
&&\theta^d\wedge\eta_{abc} = 3\delta^{d}_{[a}\eta_{bc]}\;, \qquad
\theta^f\wedge\eta_{abcd} = -4\delta^{f}_{[a}\eta_{bcd]}\;, \label{eq:eta4}
\end{eqnarray}
where $*$  is the Hodge operator. In $U_4$ the following formula is valid
\cite{Tr},
\begin{equation}
D\eta_{ab} = {\cal T}^c \wedge \eta_{abc}\;. \label{eq:eta5}
\end{equation}
\par
     Let us consider in $U_4$ the general 10-parametric Lagrangian which is
constructed from all the terms quadratic in the irreducible pieces of
curvature and torsion with the usual linear Einstein--Cartan term added. In
terms of exterior differential forms it reads \cite{Mac},
\begin{equation}
{\cal L} = f_0 {\cal R}^{a}\!_{b} \wedge
{\eta}_{a}\!^{b} + \sum_{i=1}^6 \lambda_{i} \stackrel{(i)}
{\cal R}\!^{a}\!_{b} \wedge * \stackrel{(i)}{\cal R}\!_{a}\!^{b} +
\sum_{i=1}^3 \chi_{i} \stackrel{(i)} {\cal T}\!^{a}
\wedge * \stackrel{(i)}{\cal T}_{a}\;. \label{eq:lag1}
\end{equation}
Here $f_0 = 1/(2\kappa)$ ($\kappa = 8\pi G/c^4 $), and $\lambda_{i}$,
$\chi_{i}$ are the coupling constants. The index $(i)$ runs over all
irreducible (with respect to Lorentz group) components of the curvature
2-form and the torsion 2-form, respectively. Refs. \cite{Ad}, \cite{Sin} and
some others deal with the special cases of the Lagrangian (\ref{eq:lag1}).
\par
     The torsion 2-form can be decomposed into traceless 2-form, trace 2-form
and pseudotrace 2-form as follows (see \cite{He:pr}),
\begin{equation}
{\cal T}^{a} = \stackrel{(1)}{\cal T}\!^{a} + \stackrel{(2)}{\cal T}\!^{a} +
\stackrel{(3)}{\cal T}\!^{a}\; . \label{eq:razl}
\end{equation}
Here the trace 2-form and the pseudotrace 2-form of the pseudo-Riemannian
4-manifold are determined by the expressions, respectively,
\begin{equation}
\stackrel{(2)}{\cal T}\!^{a} = \frac{1}{3} \theta^{a}\wedge (\bar e_{b}
\rfloor {\cal T}^{b})\;, \qquad \stackrel{(3)}{\cal T}\!^a = \frac{1}{3}
* (\theta^a \wedge * ({\cal T}^{b}\wedge \theta _b ))\;. \label{eq:sled}
\end{equation}
The traceless and trace parts of torsion satisfy to the conditions,
\begin{equation}
\stackrel{(1)}{\cal T}\!^{a}\wedge{\theta }_a = 0\;, \qquad
\stackrel{(2)}{\cal T}\!^{a}\wedge{\theta }_a = 0\;. \label{eq:sv1}
\end{equation}
On the decomposition of the curvature 2-form into the irreducible pieces see
Ref. \cite{He:pr}.
\par
     Using the expressions for the irreducible parts of  curvature  and
torsion 2-forms the Lagrangian (\ref{eq:lag1}) can be transformed into the
form which is more convinient for the variation procedure,
\begin{eqnarray}
&&{\cal L} = f_0 {\cal R }^{a}\!_{b}\wedge{\eta }_{a}\!^{b}
+ \tau_1 {\cal R}^{a}\!_{b}\wedge *{\cal R}^{b}\!_{a} \nonumber \\
&& + \tau_2 ({\cal R}^{ab} \wedge {\theta}_{a}) \wedge * ({\cal R}^{c}\!_{b}
\wedge {\theta}_{c}) + \tau_3 ({\cal R}^{ab}\wedge {\theta}_{c})\wedge
* ({\cal R}^{c}\!_{b}\wedge {\theta}_{a}) \nonumber \\
&& + \tau_4 ({\cal R}^{a}\!_{b}\wedge {\theta}_{a} \wedge {\theta}^{b})
\wedge * ({\cal R}^{c}\!_{d} \wedge{\theta}_{c}\wedge{\theta}^{d})\nonumber\\
&& + \tau_5 ({\cal R}^{a}\!_{b}\wedge{\theta}_{a}\wedge {\theta}^{d}) \wedge
* ({\cal R}^{c}\!_{d}\wedge{\theta}_{c}\wedge{\theta}^{b}) \label{eq:lag} \\
&& + \tau_6 ({\cal R}^{a}\!_{b}\wedge{\theta}_{c}\wedge{\theta}^{d})\wedge
* ({\cal R}^{c}\!_{d}\wedge{\theta}_{a}\wedge {\theta}^{b}) + \varrho_1
{\cal T}^{a} \wedge *  {\cal T}_{a} \nonumber \\
&& + \varrho_2 ({\cal T}^{a}\wedge {\theta}_{a}) \wedge * ({\cal T}^{b}
\wedge {\theta}_{b}) + \varrho_3 ({\cal T}^{a} \wedge {\theta}_{b}) \wedge
* ({\cal T}^{b}\wedge{\theta}_{a})\;. \nonumber
\end {eqnarray}
Here $\tau_1,...,\tau_6$,  $\varrho_1,...,\varrho_3$  are the coupling
constants. They are related to the coupling constants of the Lagrangian
(\ref{eq:lag1}) as follows:
\begin{eqnarray}
&& \tau_1 = \frac{1}{2}(3\lambda_4 - 5\lambda_1)\;,\qquad\tau_2 = \frac{1}{2}
(\lambda_2 + \lambda_4 + \lambda_5 - 3\lambda_1)\;, \qquad \tau_3  =
\frac{1}{2}(\lambda_4 - \lambda_1)\;, \nonumber \\
&& \tau_4 = \frac{1}{2}(- \lambda_2 + \lambda_3) \;, \quad \tau_5 =
\frac{1}{2} (\lambda_2 + \lambda_4 - \lambda_5 - \lambda_1)\;, \quad
\tau_6 = \frac{1}{12}(3\lambda_4 - 2\lambda_1 + \lambda_6)\;, \nonumber \\
&& \varrho_1 = \frac{1}{2}(2\chi_1 + \chi_2)\;, \qquad \varrho_2 = \frac{1}{3}
(-\chi_1 + \chi_3)\;, \qquad \varrho_3 = \frac{1}{3}(\chi_1 - \chi_2)\;,
\nonumber \end{eqnarray}
One can easily express the parameters of (\ref{eq:lag1}) through the
parameters of the Lagrangian (\ref{eq:lag}). We need only one of them, namely,
\begin{equation}
\chi_1 = \varrho_1 + \varrho_3 \; . \label{eq:chi}
\end{equation}
\par
     Beeing written in the component form, the Lagrangian (\ref{eq:lag})
coinsides with the well-known Lagrangian \cite{Hay}, ${\cal L} = L\eta$,
where
\begin{eqnarray}
&&L = f_{0}R + R^{abcd}(f_{1}R_{abcd} + f_{2}R_{acbd} +
f_{3}R_{cdab}) \nonumber \\
&& + R^{ab}(f_{4}R_{ab} + f_{5}R_{ba}) + f_{6}R^{2}+ T^{abc}(a_{1}T_{abc}
+ a_{2}T_{cba}) + a_{3}T_{a}T^{a}\; . \label{eq:Ha}
\end{eqnarray}
Here we use the following notations:
\begin{equation}
R_{ab} = R^{c}\!_{acb}\;,\qquad R = R^{c}\!_{c}\;,\qquad T_{a} =
T^{c}\!_{ac}\; . \label{eq:ri}
\end{equation}
The constants in the Lagrangian (\ref{eq:Ha}) are related to the parameters of
the Lagrangian (\ref{eq:lag}) as follows,
\begin{eqnarray}
&&f_1 = \frac{1}{2}(-\tau_1 + \tau_2 - \tau_3 + 2\tau_4 + \tau_5 +  2\tau_6)
\;, \qquad f_2 = \tau_4\;, \nonumber \\
&&f_3 = -2\tau_2 - 4\tau_4 - \tau_5\;,\quad f_4 = -\tau_3 - \tau_5 - 4\tau_6
\;, \quad f_5 = \tau_5\;, \quad f_6 = \tau_6\;, \nonumber \\
&& a_1 = \frac{1}{2}(\varrho_1 + \varrho_3 + \varrho_3)\;, \qquad
a_2 = -\varrho_2\;, \qquad a_3 = -\varrho_3\;. \nonumber
\end{eqnarray}
\par
     The vacuum field equations in the Riemann--Cartan space-time can be
obtained by the variational procedure of the first order. Let us vary the
Lagrangian (\ref{eq:lag}) with respect to the basis 1-form $\theta^{a}$ and
to the connection 1-form $\Gamma^{a}\!_{b}$ in  $U_4$ independently and take
into account that the connection 1-form satisfies to the condition,
$\Gamma_{ab} = - \Gamma_{ba}$. It is useful to use the formula (\ref{eq:var})
of Apendix. The variation with respect to $\theta^{a}$ gives the field
equation,
\begin{eqnarray}
&&f_0 {\cal R}^{bc} \wedge {\eta }_{abc} + {\tau }_1
({\cal R}^{n}\!_{b} \wedge * ( {\cal R}^{b}\!_{n} \wedge {\theta }_a )
+ * (* {\cal R}^{b}\!_{n} \wedge {\theta }_a ) \wedge * {\cal R }^{n}\!_{b})
\nonumber \\
&&+ {\tau }_2 (2 {\cal R}_{an} \wedge * ({\cal R }^{mn} \wedge{\theta }_{m})
- ({\cal R}^{rb} \wedge {\theta }_{r}) \wedge * ({\cal R}^{m }\!_{b} \wedge
{\theta }_{m} \wedge {\theta }_a ) \nonumber \\
&& - * (* ({\cal R}^{m}\!_{b} \wedge{\theta }_{m}) \wedge {\theta }_a )
\wedge * ({\cal R}^{rb} \wedge {\theta }_{r})) \nonumber \\
&&+ {\tau }_3 (2{\cal R}^{t}\!_{n}\wedge * ({\cal R}^{n}\!_{a} \wedge
{\theta }_{t}) - ({\cal R}_{l}\!^{b} \wedge {\theta }_{m})\wedge
* ({\cal R }^{m}\!_{b} \wedge {\theta }^{l} \wedge {\theta }_{a})\nonumber \\
&&- * (* ({\cal R}^{m}\!_{b}\wedge {\theta }^{\lambda })\wedge {\theta }_a )
\wedge * ({\cal R}_{l}\!^{\beta }\wedge {\theta }_{m}) ) \nonumber \\
&&+ {\tau }_{4} (4{\cal R}_{ab}\wedge {\theta }^{b}\wedge{\theta }^{b}\wedge
* ({\cal R}^{ln}\wedge{\theta }_{l}\wedge{\theta }_{n}) + * (* ({\cal R}^{l}
\!_{n}\wedge{\theta }_{l}\wedge {\theta }^{n}) \wedge {\theta }_a )\wedge
* ({\cal R}^{t}\!_{b} \wedge {\theta }_{t}\wedge {\theta }^{b}) )\nonumber \\
&&+ {\tau }_{5} (4{\cal R}_{[a|b|}\wedge {\theta }^{l}\wedge * ({\cal R}^{n}
\!_{l]}\wedge{\theta }_{n}\wedge{\theta }^{b}) + * (* ({\cal R}^{n}\!_{l}
\wedge{\theta }_{n}\wedge{\theta }^{b})\wedge {\theta }_a )\wedge * (
{\cal R}^{t}\!_{b} \wedge {\theta }_{t}\wedge {\theta }^{l}) ) \nonumber \\
&&+ {\tau }_{6} (4{\cal R }^{n}\!_{b} \wedge {\theta }^{l} \wedge *
({\cal R }_{al}\wedge{\theta }_{n}\wedge{\theta }^{b}) + * ( * ({\cal R}^{l}
\!_{n}\wedge{\theta }^{t}\wedge {\theta }^{b})\wedge {\theta }_a ) \wedge *
({\cal R}_{tb}\wedge{\theta }_{l}\wedge {\theta }^{n} ) ) \nonumber \\
&&+ {\varrho }_1 (2D*{\cal T}_a + {\cal T}_n\wedge * ({\cal T}^n\wedge {\theta }
_a ) + * (* {\cal T}^n\wedge {\theta }_a )\wedge * {\cal T}_n ) \nonumber \\
&&+ {\varrho }_2 (4{\cal T}_a\wedge * ({\cal T}^{b}\wedge {\theta }_{b}) - 2
{\theta }_a\wedge D* ({\cal T}^b \wedge {\theta }_b ) - ({\cal T}^{l}\wedge
{\theta }_{l}) \wedge * ({\cal T}^b\wedge{\theta }_b \wedge {\theta }_a ) )
\nonumber \\
&&+ {\varrho }_2 (* (* ({\cal T}^b\wedge{\theta }_b )\wedge {\theta }_a )
\wedge * ({\cal T }^l \wedge {\theta }_{l}) ) \nonumber \\
&&+ {\varrho }_3 (2{\cal T}^{b }\wedge * ({\cal T}_a \wedge {\theta }_b ) +
2 D ({\theta }^b\wedge * ({\cal T}_b\wedge {\theta }_a ) ) - ({\cal T}_l\wedge
{\theta }_n )\wedge * ({\cal T}^n \wedge {\theta }^l \wedge {\theta }_a ) )
\nonumber \\
&&+ {\varrho }_3 (* (* ({\cal T}^n\wedge{\theta }^l )\wedge {\theta }_a )\wedge
* ({\cal T}_l\wedge {\theta }_n ) ) = 0 \;. \label{eq:vart}
\end{eqnarray}
The second field equation is the result of the variation with respect to
$\Gamma^{a}\!_{b}$ and has the form,
\begin{eqnarray}
&& f_0 D \eta_{a}\!^{b} + \tau_{1}2 D*{\cal R}^{b}\!_{a} + \tau_{2}D (
{\theta}_{a}\wedge * ({\cal R}^{cb}\wedge{\theta}_{c}) - {\theta}^{b}\wedge
* ({\cal R}^{c}\!_{a}\wedge {\theta}_{c}) ) \nonumber \\
&& + \tau_{3} D ({\theta }^{c}\wedge * ({\cal R}_{c}\!^{b}\wedge{\theta}_{a})
- {\theta}^{c}\wedge * ({\cal R}_{ca}\wedge {\theta}^{b}) ) \nonumber \\
&& + \tau_{4}2 D ({\theta}_{a}\wedge {\theta}^{b}\wedge * ({\cal R}^{cd}
\wedge{\theta}_{c}\wedge{\theta}_{d}) ) \nonumber \\
&& + \tau_{5} D ({\theta}_{a}\wedge {\theta}^{d}\wedge * ({\cal R}^{c}\!_{d}
\wedge {\theta}_{c}\wedge{\theta}^{b}) - {\theta}^{b}\wedge{\theta}^{d}\wedge
* ({\cal R}^{c}\!_{d}\wedge{\theta}_{c}\wedge {\theta}_{a}) ) \nonumber \\
&& +\tau_{6}2 D ({\theta}_{c}\wedge{\theta}^{d} \wedge * ({\cal R}^{c}\!_{d}
\wedge{\theta}_{a}\wedge {\theta}^{b}) ) \nonumber \\
&& + \varrho_{1} ({\theta}^{b}\wedge * {\cal T}_{a} - {\theta}_{a}\wedge *
{\cal T}^{b}) + \varrho_{2} 2 {\theta}^{b}\wedge{\theta}_{a}\wedge * (
{\cal T}^{c}\wedge{\theta}_{c}) \nonumber \\
&& + \varrho_{3} ({\theta}^{b}\wedge{\theta}^{c}\wedge * ({\cal T}_{c}\wedge
{\theta}_{a}) - {\theta}_{a}\wedge{\theta}^{c}\wedge * ({\cal T}_{c}\wedge
{\theta}^{b}) ) = 0\;. \label{eq:varg}
\end{eqnarray}

\section{\bf Plane torsion waves in Riemann--Cartan space-time}
\setcounter{equation}{0}
\par
     As it is shown in \cite{EK}, \cite{4}, it is convinient to choose the
special basis to investigate the problem of gravitational waves. This basis
is constructed from two null vectors $\bar e_{0}=\partial_{v}$, $\bar e_{1}=
\partial_{u}$ and two space-like vectors $\bar e_{3} = \partial_{x}$,
$\bar e_{4} = \partial_{y}$. The vector $\bar e_{0}$ is covariant constant
and  has the direction of the wave ray. Coordinates $x$ and $y$ parametrize
the wave surface $(u,\,v) = \mbox{const}$. In this basis the metric tensor
has the form,
\begin{displaymath}
g_{ab}=
\left(
\begin{array}{cccc}
0&1&0&0\\
1&0&0&0\\
0&0&-1&0\\
0&0&0&-1
\end{array}\right)\;.
\end{displaymath}
\par
     In a Riemann space $V_{4}$ the plane wave of a metric (see Refs. \cite{Ad},
\cite{4}) is defined as the special case of the metric for the plane-fronted
gravitational waves with parallel rays (pp-waves), which in the basis chosen
has the form,
\begin{equation}
g = 2 H(x,y,u) du^2 + 2 dudv - dx^2 - dy^2\;. \label{eq:met}
\end{equation}
Here the coordinate $u$ is considered as the retarded time parameter and can
be interpreted as the phase of the wave. The vacuum Einstein equations for
the metric (\ref{eq:met}) are equivalent to the equation $H_{xx}+H_{yy} = 0$,
where $H_{xx}$, $H_{yy}$ are the second partial derivatives of $H$ with
respect to the corresponding coordinates. The following null coframe
corresponds to the metric (\ref{eq:met}),
\begin{equation}
\theta^0 = H du + dv\;, \quad \theta^1 = du\;, \quad \theta^2 = dx\;, \quad
\theta^3 = dy\;. \label{eq:tet}
\end{equation}
The Riemann space $V_{4}$ with the metric of the plane wave admits the
$G_5$ group of symmetries. This group is generated by the vector fields $X$
with the structure \cite{Ad},
\begin{equation}
X = (a + b'x + c'y)\partial _v + b\partial _x + c\partial _y \;, \label{eq:x}
\end{equation}
where $a = const$,  $b(u)$, $c(u)$ are arbitrary functions and $b'$, $c'$
are their derivatives with respect to $u$. The $G_5$-group of symmetries leaves
the isotropic hypersurfaces in  $V_4$ to be unchanged. These hypersurfaces
describe the wave fronts with the constant amplitude. As the special case of
pp-waves, the plane waves have the  vanishing expansion, twist and they are
shear-free \cite{Zakh}.
\par
     In Ref. \cite{Ad} the definition of the plane wave metric is extended to
the space with nonvanishing torsion. Let us introduce the following
definition \cite{India}, \cite{Izv}. \newline
{\em Definition.}
We shall call a Riemann--Cartan space $U_4$ as a space $U_4$ of a plane wave
type and its metric and torsion as the plane waves of a metric and torsion,
if the metric $g_{ab}$ and the torsion 2-form ${\cal T }^{a}$ of this space
admit a five-dimentional group  $G_5$ of symmetries. It means that the
following conditions are fullfilled: $\mbox L_{X}{g_{ab}} = 0$, $\mbox L_{X}
{\cal T}^{a} = 0$, where $\mbox L_{X}$ denotes the Lie derivative with
respect to any vector field $X$  which generates the $G_5$ symmetry.
\par
     The following theorem determines the structure of the plane torsion
waves. \newline\newline
{\em Theorem 3.1.} The torsion 2-form of $U_4$ of the plane
wave type  has the following structure:  its trace and pseudotrace
vanish and its traceless part depends on two arbitrary functions
$t$ and $s$ of the retarded time parameter $u$ and has the form,
\begin{equation}
\stackrel{(1)}{\cal T}\!^{0} = t(u){\theta }^1 \wedge {\theta }^2 + s(u)
{\theta }^1 \wedge {\theta }^3\; . \label{eq:tor}
\end{equation}
{\em Proof.}
Let us substitute the vector field (\ref{eq:x}) into the equation
$\mbox{L}_{X}{\cal T}^{a} = 0$. As a result one obtains the system of the
equations. Since $b$, $c$, $b'$ and $c'$ can take arbitrary values, these
equations imply that all the components of torsion vanish except
$T^{0}\!_{12}$ and $T^{0}\!_{13}$, which are the functions of $u$. The
straightforward verification shows that the trace and pseudotrace of torsion
(\ref{eq:sled}) of plane waves vanish and only the
traceless part of torsion depends on these nonvanishing components, and the
dependence has the form (\ref {eq:tor}).
\par
     The following theorem specifies the conditions under which the plane
torsion waves exist. \newline \newline
{\em Theorem 3.2.} The metric and torsion of $U_4$ of plane wave type are the
solutions of the field equations of the gravitational theory with the
Lagrangian (\ref{eq:lag}) with $f_0 \not = 0$ if and only if \enspace (a)
the  metric of $U_4$ satisfies the equation $\stackrel{R}R\!_{ab} = 0$, where
$\stackrel{R}R\!_{ab}$ is the Ricci tensor of the Riemann connection;
\enspace (b) the following restriction on the parameters of the Lagrangian
(\ref{eq:lag}) is valid, $f_0 + \chi_{1} = 0$. If the condition (b)
is not satisfied then torsion of the plane wave vanishes.
\newline {\em Proof.} The connection 1-form of a Riemann--Cartan space $U_4$
can be decomposed into a sum of the Riemann (Levi-Chivita) connection 1-form
$\stackrel{R}{\Gamma}\!^{a}\!_{b}$ and the contorsion 1-form
${\cal K}^{a}\!_{b}$.  The latter lineary depends on torsion \cite{He:pr},
\begin{eqnarray}
\Gamma^{a}\!_{b} = \stackrel{R}{\Gamma}\!^{a}\!_{b} + {\cal K}^{a}\!_{b}\;,
\quad {\cal T}^{a} =: {\cal K}^{a}\!_{b}\wedge\theta^{b}\;, \quad
{\cal K}_{ab} = 2\bar{e}_{ [a}\rfloor {\cal T}_{b]} - \frac{1}{2}\bar{e}_{a}
\rfloor\bar{e}_{b}\rfloor ({\cal T}_{c}\wedge\theta^{c})\;. \label{eq:kon2}
\end{eqnarray}
In consequence of (\ref{eq:kon2}) the decomposition of the curvature 2-form
in $U_4$ into the Riemannian and post-Riemannian parts reads,
\begin{equation}
{\cal R}^{a}\!_{b} = \stackrel{R}{{\cal R}}\!^{a}\!_{b} + \stackrel{R}
{D}{\cal K}^{a}\!_{b} + {\cal K}^{a}\!_{c}\wedge{\cal K}^{c}\!_{b}\;,
\label{eq:kriv}
\end{equation}
where $\stackrel{R}{D}$ is the external covariant derivative with respect to
the Riemann connection 1-form, and $\stackrel{R}{{\cal R}}\!^{\alpha}\!_
{\beta}$ is the Riemann curvature 2-form. By the explicit calculation using
(\ref{eq:tet}), (\ref{eq:tor}) and (\ref{eq:kon2}) one can verify that all
nonvanishing components of the Riemann connection 1-form and the contorsion
1-form are proportional to the basis form $\theta^1 = du$,
\begin{eqnarray}
&&\stackrel{R}{\Gamma}\!^0\!_2 = \stackrel{R}{\Gamma}\!^2\!_1 = H_x
\theta^1\;, \qquad \stackrel{R}{\Gamma}\!^0\!_3 = \stackrel{R}
{\Gamma}\!^3\!_1 = H_y \theta^1\;, \label{eq:gam}\\
&& {\cal K}^0\!_2 = {\cal K}\!^2\!_1 = t \theta^1\;,\qquad {\cal K}\!^0\!_3 =
{\cal K}\!^3\!_1 = s \theta^1\;, \label{eq:kon3}
\end{eqnarray}
where  $H_x$ and $H_y$ are the first partial derivatives of $H$ with respect to
the correspoinding coordinates. In consequence of (\ref{eq:gam}) the
external covariant derivative with respect to the Riemann connection is
also proportional to the basis 1-form $\theta^1 = du$. Therefore all the
non-Riemannian terms in (\ref{eq:kriv}) vanish and the curvature 2-form of
$U_4$ of the plane wave type coinsides with the Riemann curvature 2-form.
The only nonvanishing in the basis (\ref{eq:tet}) components of the curvature
2-form read,
\begin{equation}
{\cal R}^0\!_2 = {\cal R}^2\!_1 = H_{xx} {\theta  }^{2} \wedge {\theta}^{1}
+ H_{xy}{\theta }^{3} \wedge {\theta }^{1}\;, \quad
{\cal R}^0\!_3 = {\cal R}^3\!_1 = H_{xy} {\theta }^{2} \wedge {\theta }^{1}
+ H_{yy} {\theta }^3 \wedge {\theta }^{1}\;. \label{eq:R}
\end{equation}
Therefore in case of the plane waves of a metric and torsion the curvature
2-form ${\cal R}\!^a\!_b$ in the field equations (\ref{eq:vart}) and
(\ref{eq:varg}) coinsides with the curvature of the Riemann connection
$\stackrel{R}{{\cal R}}\!^{a}\!_b$. Then in consequence of the identity for
the Riemann curvature 2-form
\begin{equation}
\stackrel{R}{\cal R}\!^{a}\!_{b}\wedge \theta^{b} = 0\;, \label{eq:cicl}
\end{equation}
all the terms in the field equation (\ref{eq:vart}) with the coefficients
${\tau }_2$, ${\tau }_4$ and ${\tau }_5$ vanish. Moreover, as the result of
the curvature 2-form structure (\ref{eq:R}) the terms with the coefficients
${\tau }_1$, ${\tau }_3$ and ${\tau }_6$ also vanish from the equation
(\ref{eq:vart}). Taking into account that trace and pseudotrace of torsion
vanish in the case considered and that for the tracefree part the condition
(\ref{eq:sv1}) is valid, one can varify that the field equation
(\ref{eq:vart}) is equivalent to the equation,
\begin{equation}
f_0 \stackrel{R}{\cal R}\!^{bc} \wedge {\eta }_{abc} +
2 \left( \varrho_1+\varrho_3 \right) D*{\cal T}_a = 0\;. \label{eq:sss}
\end{equation}
From the structure of the torsion 2-form (\ref{eq:tor}) and the connection of
the plane wave (\ref{eq:gam}), (\ref{eq:kon3}) it follows that the last term
in (\ref{eq:sss}) vanishes. Then the equation (\ref{eq:sss}) takes the form,
\begin{equation}
f_{0} \stackrel{R}R\!_{ab} = 0\; .  \label{eq:ein}
\end{equation}
This proves the part (a) of Theorem 3.2.
\newline
Let us now consider the field equation (\ref{eq:varg}). In this equation as
well as in (\ref{eq:vart}) the curvature 2-form coinsides with the curvature
of the Riemann connection in $V_4$. Then due to (\ref{eq:cicl}) the terms
with the coefficients ${\tau }_2$, ${\tau }_4$ and ${\tau }_5$ vanish. Next,
using the component representation of the curvature 2-form, together with the
identities (\ref{eq:eta1})--(\ref{eq:eta4}), one can show that the terms with
the coefficients ${\tau }_3$ and ${\tau }_6$ are equal to
\begin{eqnarray}
&&{\tau }_3 D (2*{\cal R}_a\!^b + R_{ac}\eta^{bc} - R^{bc}\eta_{ac})\;,
\label{eq:t3}\\
&& 8 {\tau }_6 D (*{\cal R}_a\!^b + R_{ac}\eta^{bc} - R^{bc}\eta_{ac} +
\frac{1}{4} R \eta_a\!^b)\;, \label{eq:t6}
\end{eqnarray}
where the notations (\ref{eq:ri}) are used. It is necessary to take into
account that because of the equation (\ref{eq:ein}) the Ricci tensor and the
scalar curvature vanish in (\ref{eq:t3}) and (\ref{eq:t6}). The term with the
coefficient $f_0$ in (\ref{eq:varg}) can be transformed with the help of
(\ref{eq:eta5}). Now let us take into account the fact that the trace and
pseudotrace of the torsion vanish and also use the following identities for
the traceless part of torsion,
\begin{equation}
*\stackrel{(1)}{\cal T}\!^{b} = \theta^c \wedge *(\stackrel{(1)}{\cal T}\!_{c}
\wedge\theta^{b})\;, \quad
\stackrel{(1)}{\cal T}\!^{c}\wedge\eta^{b}\!_{ac} = \theta_a \wedge
*\stackrel{(1)}{\cal T}\!^{b}  -  \theta^b \wedge *\stackrel{(1)}
{\cal T}\!_{a}\;. \label{eq:id}
\end{equation}
The identities (\ref{eq:id}) can be verified by choosing the component
representation and using (\ref{eq:eta1})--(\ref{eq:eta4}). Finally, in the
equation obtained let us decompose the connection 1-form into its Riemannian
and post-Riemannian parts according to (\ref{eq:kon2}). As a result the
equation (\ref{eq:varg}) for the Riemann--Cartan space of the plane wave type
becomes,
\begin{eqnarray}
&& {\tau }\stackrel{R}{D}*\stackrel{R}{{\cal R }}\!^{b}\!_{a} + 2\tau \left(
(\bar e^{[b}\rfloor\stackrel{(1)}{\cal T}\!_{c]})\wedge *\stackrel{R}{{\cal R}}
\!^{c}\!_{a} - (\bar e^{[c}\rfloor \stackrel{(1)}{\cal T}\!_{a]})\wedge
*\stackrel{R}{{\cal R }}\!^{b}\!_{c}\right) \nonumber \\
&& + (f_0 + \varrho_{1} + \varrho_3)\left (*\stackrel{(1)}{\cal T}\!_{a}
\wedge {\theta }^b -*\stackrel{(1)}{\cal T}\!^{b}\wedge {\theta }_a\right)
= 0\; , \label{eq:urg1}
\end{eqnarray}
where $\tau = 2(\tau_1 - \tau_3 - 4\tau_6)$. Let us consider each term of
this equation separately. The first term vanishes by virtue of the Bianchi
identity in $V_4$ and the equation (\ref{eq:ein}). The second term vanishes
due to the structure of the curvature 2-form (\ref{eq:R}) and the torsion
2-form (\ref{eq:tor}). As a result the equation (\ref{eq:urg1}) reduces to
the system of the equations,
\begin{equation}
(f_0 + \varrho_{1} + \varrho_3 ) t(u) = 0\;, \qquad
(f_0 + \varrho_{1} + \varrho_3 ) s(u) = 0\;. \label{eq:urt}
\end{equation}
The rest of the statements of Theorem 3.2 follow from (\ref{eq:urt})
due to (\ref{eq:chi}) and (\ref{eq:tor}).\newline
{\em Corollary.}
Only the traceless part of torsion can propagate in the form of a plane wave.
Traceless torsion plane waves have the massless quanta.\newline
{\em Proof.} Using (\ref{eq:razl}) one can decompose the linear term of the
Lagrangian (\ref{eq:lag1}) into the Riemannian and post-Riemannian parts as
follows,
\begin{eqnarray}
&&{\cal R}^{a}\!_{b}\wedge \eta_{a}\!^{b} = \stackrel{R}{{\cal R}}\!^{a}\!_{b}
\wedge\eta_{a}\!^{b} - 2 \stackrel{R}{D} \left (*\stackrel{(2)}{\cal T}\!^{a}
\wedge\theta_{a}\right)  \nonumber \\
&& + \stackrel{(1)}{\cal T}\!^{a} \wedge * \stackrel{(1)}{\cal T}\!_{a}
- 2\stackrel{(2)}{\cal T}\!^{a}\wedge * \stackrel{(2)}{\cal T}\!_{a}
- \frac{1}{2}\stackrel{(3)}{\cal T}\!^{a}\wedge
*\stackrel{(3)}{\cal T}\!_{a}\; . \label{eq:bb}
\end{eqnarray}
Using (\ref{eq:bb}) one obtains that the coefficient at the quadratic
traceless part of torsion in (\ref{eq:lag1}) is $f_0+\chi_{1}$ and is equal
to zero in consequence of Theorem 3.2. This can be interpreted as the fact
that the quanta of the corresponding part of torsion are massless.

\section{\bf Discussion}
\setcounter{equation}{0}
     In this article the gravitational field equations for the general
10-parametric quadratic  Lagrangian  are  obtained  in   the   first   order
variational formalism in $U_4$ in terms of the exterior differential forms.
The variational procedure is based on the formula for the commutator of
variation of the arbitrary p-form and the Hodge operation, which is proved
as the Lemma in the mathematical Appendix. The field equations obtained are
used for the analysis of the problem of the plane torsion waves in the
quadratic gravitational theories.
\par
      For this purpose the definition of the plane waves of a metric and
torsion as the Riemann--Cartan space of the plane wave type is given. The
structure of the torsion 2-form in the case of the plane waves  is determined
by Theorem 3.1. From this Theorem it follows that only the traceless part
of torsion can propagate in the form of the plane wave. In connection with
this result the important question of the possible sources of such waves
arises, since the usual types of matter (spinor and electromagnetic fields)
can not be the source of the torsion traceless part.
\par
     The necessary and sufficient conditions for the existense of the plane
torsion waves are found. They are determined by Theorem 3.2. These
conditions consist of two parts. The first one requiers the metric of the
plane wave to be the plane wave metric with respect to the connection of the
Riemann space $V_4$, whose metric coinsides with the  metric of the
Riemann--Cartan space $U_4$. The second condition requiers the constraint to
be imposed on the parameters of the quadratic Lagrangian, $f_0 + \chi_{1} =
0$. As follows from Corollary of Theorem 3.2 the last condition leads to the
fact that the traceless part of torsion must have the massless quanta.
\par
     The restriction $f_0 + \chi_{1} = 0$ on the Lagrangian parameters plays
the important role in analysis of various quadratic gravitational theories
with torsion. Thus, in \cite{Zang} it is shown that if this condition is
fulfilled, the spherically symmetric solution for the torsion has the
$1/r^2$ form. This leads to the theory without tachyons on the quantum level.
In \cite{Sez} five types of the Lagrangian which lead to the theory without
ghosts and tachyons are found. Among them there are two Lagrangians for which
this restriction on the constants holds. This leads in turn to the possible
propagation of the plane torsion waves. The condition $f_0 + \chi_{1} = 0$
is also important in the investigation of particle spectrum in the Poincar\'e
gauge gravitational theory. In \cite{Hay-Shi} it is shown that if this
condition is valid then the normal multiplet of torsion can be  constructed
only from the following combinations of the irreducible components:
$\left( 2^{-}, 1^{-}, 0^{+} \right)$,  $\left( 2^{-}, 1^{+}, 0^{+} \right)$,
$\left( 2^{-}, 0^{+}, 0^{-} \right)$.

\section*{Appendix}
\renewcommand{\theequation}{A.\arabic{equation}}
\setcounter{equation}{0}
     We consider the arbitrary (in general anholonimic) basis
$\bar{e}_{\alpha}$. For this basis the cobasis of 1-forms $\theta^\beta$
does not consist of total differentials. The metric takes the form
$g = g_{\alpha\beta}\theta^\alpha \otimes \theta^\beta$, where
$g_{\alpha\beta} = g(\bar{e}_\alpha ,\,\bar{e}_\beta )$. The following
Lemma is valid. \newline \newline
{\em Lemma.} Let $\Phi$ and $\Psi$ be arbitrary $p$-forms defined on
$n$-dimentional manifold. Then the following variational identity for the
commutator of the variation operator and the Hodge star operator is valid,
\begin{eqnarray}
&& \Phi \wedge \delta * \Psi = \delta \Psi \wedge * \Phi + \delta
g_{\sigma \rho } \left( \frac{1}{2} g^{\sigma \rho } \Phi
\wedge * \Psi + (-1)^
{p(n-p)+\mbox{\small{Ind}}(g) + 1}
{\theta }^{\sigma } \wedge
* \left( * \Psi \wedge {\theta }^{\rho } \right) \wedge * \Phi \right)
\nonumber \\
&&+ \delta {\theta }^{\alpha } \wedge \left( {(-1)}^p \Phi \wedge
* \left( \Psi \wedge {\theta }_{\alpha } \right)  +
{(-1)}^{p(n-p)+\mbox{\small{Ind}}(g) + 1}*\left ( * \Psi \wedge {\theta }_{\alpha }
\right) \wedge * \Phi \right)\; . \label{eq:var}
\end{eqnarray}
Here $\delta$ is the operator of the variational derivative and
$\mbox{Ind}(g)$ is the index of the metric $g$, which is equal to the number
of negative eigenvalues of the diagonalized metric.\newline
{\em Proof.} Let $\Psi$ be an arbitrary $p$-form with the  component
representation
\begin{equation}
\Psi = \frac{1}{p!} {\Psi }_{\gamma _1\gamma _2...\gamma _p}
\theta^{\gamma_1}\theta^{\gamma_2}...\theta^{\gamma_p}\; . \label{eq:psi}
\end{equation}
Then the Hodge star operator applied to this $p$-form reads,
\begin{equation}
*\Psi = \frac{\sqrt{|\mbox{det}g_{\sigma\rho}|}}{(n-p)!p!}g^{\alpha_1
\gamma_1}...g^{\alpha_p\gamma_p}\epsilon_{\alpha_1...\alpha_p\beta_1...\beta
_{n-p}}\Psi_{\gamma_1...\gamma_p}
\theta^{\beta_1}\wedge ...\wedge \theta^{\beta_{n-p}}\;, \label{eq:Hodg}
\end{equation}
where $\epsilon_{\alpha_1...\alpha_p\beta_1...\beta_{n-p}}$ are the
components of the totally antysymmetric Levi--Chivita density $n$-form.
Let us calculate the variation $\delta\Psi $ using (\ref{eq:psi}). Then
applying (\ref{eq:Hodg}) one gets,
\begin{eqnarray}
&&*\delta\Psi = \frac{1}{p!}{\eta }^{{\gamma }_1...{\gamma }_p}\delta {\Psi }
_{{\gamma }_1...{\gamma }_p} + \frac{1}{(p-1)!}{\Psi }_{{\gamma }_1...
{\gamma }_p} \left (\bar{e}_{\lambda } \rfloor \delta {\theta }^{{\gamma }_1}
\right ) {\eta }^{\lambda {\gamma }_2...{\gamma }_p}\;. \label{eq:*d}
\end{eqnarray}
The calculation of the variation of (\ref{eq:Hodg}) is more complicated
since the components of the metric tensor must be also varied. After rather
tedious calculations one gets,
\begin{eqnarray}
&&\delta *\Psi = \delta g_{\sigma\rho }\left (\frac{1}{2}(*\Psi )g^{\sigma
\rho } - \frac{1}{(p-1)!}{\Psi }^{\sigma {\gamma }_1...{\gamma }_{p-1}}
{\eta }^{\rho}\!_{{\gamma }_1 ... {\gamma }_{p-1}} \right ) \nonumber \\
&& + \frac{1}{p!}{\eta }^{{\gamma }_1...{\gamma }_p}\delta {\Psi }_{{\gamma }
_1...{\gamma }_p} + \delta \theta^\beta \wedge \left (\frac{1}{p!}{\Psi }^
{{\gamma }_1...{\gamma }_p} {\eta }_{{\gamma }_1...{\gamma }_p \beta }
\right )\;. \label{eq:d*}
\end{eqnarray}
Here the $p$-form ${\eta }^{{\gamma }_1...{\gamma }_p} = *(\theta^
{{\gamma }_1}\wedge ...\wedge\theta^{{\gamma }_p})$ is introduced. Let us
multiply externally from the left (\ref{eq:*d}) and (\ref{eq:d*}) by the
$p$-form $\Phi$ and take into account that $\Phi\wedge *\delta \Psi =
\delta \Psi\wedge *\Phi$. Comparing the expressions obtained and using the
well-known relations,
\begin{equation}
\bar{e}_\alpha \rfloor *\Psi = * (\Psi\wedge \theta_\alpha)\;, \qquad
**\Psi = (-1)^{p(n-p) + \mbox{\small{Ind}}(g)}\Psi\;,\nonumber
\end{equation}
one gets the formula (\ref{eq:var}) of Lemma.

\end{document}